\documentclass[conference]{IEEEtran}
\IEEEoverridecommandlockouts
\usepackage{cite}
\usepackage{amsmath,amssymb,amsfonts}
\usepackage{algorithmic}
\usepackage{textcomp}
\usepackage{xcolor}
\usepackage{todonotes}
\setuptodonotes{inline}
\usepackage{lipsum}
\usepackage{float}
\usepackage{makecell}

\def\BibTeX{{\rm B\kern-.05em{\sc i\kern-.025em b}\kern-.08em
    T\kern-.1667em\lower.7ex\hbox{E}\kern-.125emX}}

\usepackage{graphicx}

\usepackage[english]{babel}
\usepackage{booktabs}  
\usepackage{multirow}  
\usepackage{hyperref}  

\begin{document}

\title{HiSTM: Hierarchical Spatiotemporal Mamba\\for Cellular Traffic Forecasting}

\author{\IEEEauthorblockN{Zineddine Bettouche, Khalid Ali, Andreas Fischer, Andreas Kassler}
    \IEEEauthorblockA{Deggendorf Institute of Technology \\
        Dieter-Görlitz-Platz 1, 94469 Deggendorf \\
        \{zineddine.bettouche, khalid.ali, andreas.fischer, andreas.kassler\}@th-deg.de}
}

\maketitle

\begin{abstract}
    Cellular traffic forecasting is essential for network planning, resource allocation, or load-balancing traffic across cells. However, accurate forecasting is difficult due to intricate spatial and temporal patterns that exist due to the mobility of users. Existing AI-based traffic forecasting models often trade-off accuracy and computational efficiency. We present Hierarchical SpatioTemporal Mamba (HiSTM), which combines a dual spatial encoder with a Mamba-based temporal module and attention mechanism. HiSTM employs selective state space methods to capture spatial and temporal patterns in network traffic. In our evaluation, we use a real-world dataset to compare HiSTM against several baselines, showing a 29.4\% MAE improvement over the STN baseline while using 94\% fewer parameters. We show that the HiSTM generalizes well across different datasets and improves in accuracy over longer time-horizons.
    
\end{abstract}


\section{Introduction}
Accurate traffic forecasting plays a critical role in enabling predictive network resource allocation, network planning and network optimization, directly influencing the operational expenditure of telecom providers~\cite{shao2020spatial}. Traditional time series forecasting methods such as ARIMA~\cite{box2015time} and AI-based models such as LSTM~\cite{hochreiter1997long} often treat each base station independently and fail to account for spatial dependencies among neighboring cells. To address this challenge, recent work focused on employing AI-based models jointly with data pre-processing and feature engineering techniques to incorporate spatial knowledge in the training procedure~\cite{hachemi2020towards,wang2020gaussian}. 

A key challenge in spatiotemporal modeling is the trade-off between using a single global model and multiple cell-specific models. While cell-specific models can capture local dynamics with higher fidelity, they are costly to train (as they require one model per cell), validate, deploy, and maintain, especially at the scale of modern cellular networks. Conversely, global models may benefit from exploiting shared patterns across cells but often underperform due to distributional heterogeneity~\cite{shao2020spatial}.

Alternatively, a global model can be trained to cover the entire spatial grid while incorporating the spatial dependencies as features, formulating the forecasting task as a multivariate timeseries forecasting problem. However, such global models often marginally improve performance due to their limited ability to capture the intricate spatiotemporal dynamics effectively. In 5G networks, user mobility and overlapping coverage areas introduce non-trivial spatial correlations~\cite{yu2021spatiotemporal}, making standalone temporal modeling insufficient as they may introduce large forecasting errors. 

Consequently, a single spatiotemporal model capturing both temporal dynamics and latent spatial interactions may exploit correlations among neighboring cells better, leading to increased forecasting accuracy. Indeed, 
recent advancements using graph neural networks (GNNs) and transformers have shown promise in learning complex spatiotemporal dependencies~\cite{gu2021glsttn}. However, these architectures typically suffer from high computational cost, making them less viable for real-time or large-scale deployments.

To address these challenges, we propose  HiSTM, a novel spatiotemporal forecasting model based on the Mamba architecture~\cite{gu2024mamba},  leveraging a hierarchical structure that captures both local variability and global patterns. Unlike graph-based or attention-heavy models, HiSTM demonstrates promising performance with reduced resource requirements, offering a favorable trade-off between accuracy and computational efficiency. Our contributions are summarized as follows:
\begin{itemize}
    \item We propose HiSTM, a novel spatiotemporal forecasting architecture that leverages the standard Mamba model combined with optimized data processing—applying spatial CNNs per frame, temporal modeling over spatial patches via Mamba, and temporal attention to aggregate time-step features for improved representation.
    \item We apply HiSTM on real-world open-source cellular traffic datasets, demonstrating improved forecasting accuracy (Milan dataset) and consistent generalization ability when evaluated on unseen dataset of a different city (Trentino dataset).
    \item We show that HiSTM achieves a 29.4\% MAE reduction compared to the STN baseline and competitive performance against other methods.
    \item HiSTM achieves better accuracy over longer forecasting horizons, with a 58\% slower error accumulation rate than STN, highlighting its ability to capture long-term temporal dependencies.
\end{itemize}

The rest of this paper is structured as follows: Section II reviews  5G traffic prediction methods. Section III formalizes the forecasting problem, introduces HiSTM and baselines we compare against. Section IV describes the datasets and experimental setup. Section V presents results and analysis. Section VI concludes the study and outlines directions for future work.

\section{Related Work}\label{relatedWork}

Traditionally, 5G traffic forecasting is  treated  as a purely temporal forecasting task, relying on recurrent neural networks. AI-based models such as vanilla RNNs, LSTMs~\cite{hochreiter1997long}, and GRUs~\cite{cho2014learning} were applied to learn sequential traffic patterns, often achieving gains over classical time-series methods such as ARIMA~\cite{box2015time}. However, these approaches inherently ignore spatial correlations. Prior studies note that while RNN-based methods can capture long-term dependencies, they neglect spatial context, reducing their effectiveness~\cite{ma2019traffic}\cite{li2021survey}. For example, Chen et al. proposed a multivariate LSTM (MuLSTM) using dual streams for traffic and handover sequences, but the model remains fundamentally temporal~\cite{chen2021data}. Temporal models employ hybrid or preprocessing strategies for improved accuracy. For instance, Hachemi et al.~\cite{hachemi2020towards} applied an FFT filter before LSTM to separate periodic signals, while Wang and Zhang~\cite{wang2020gaussian} used Gaussian Process Regression alongside LSTM to manage bursty patterns. These methods often operate at hourly aggregation levels and improve accuracy in long-term forecasts, though they still lack spatial awareness and incur higher computational cost.

Recently, spatial models have become popular as they can exploit spatial correlations within the data. For example, grid-based methods apply ConvLSTM~\cite{shi2015convolutional} or 3D-CNNs to traffic maps~\cite{zhang2019long}, though these require regular cell layouts. Graph-based architectures, such as STCNet and its multi-component variant A-MCSTCNet, combine CNNs or GRUs with attention mechanisms~\cite{narmanlioglu2022multi}.  STGCN-HO uses graph convolutions based on handover-derived adjacency to model spatial links across base stations, with Gated Linear Units (GLUs) for time dependencies~\cite{borcea2023stgcn}. Many of these models supplement deep networks with auxiliary features, such as weekday labels or slice-specific parameters, to improve generalization. More advanced spatiotemporal architectures, such as DSTL, adopt a dual-step transfer learning scheme that clusters gNodeBs and fine-tunes shared RNN models per group~\cite{aziz2025dstl}. It forecasts at a 10-minute interval while reducing training overhead. To the best of our knowledge, Mehrabian et al.~\cite{mehrabian2025agamba} published the first model to incorporate the Mamba framework into a spatiotemporal graph-based predictor for 5G traffic. It adapts a dynamic graph structure within a bidirectional Mamba block to model complex spatial and temporal dependencies. 

In contrast, our model, HiSTM, uses the Mamba framework and adapts data input and output processing to work directly on spatial grids with temporal attention layers. This design improves accuracy and efficiency, especially for long-term forecasts and generalization to unseen data.


\section{Methodology}\label{methodology}

This section formalizes the spatiotemporal forecasting problem, describes the proposed model architectures, and outlines the baseline models that we compare against.


\subsection{Problem Formulation}

Cellular network traffic forecasting aims to predict future traffic volume based on historical observations. We formulate this task as a spatiotemporal sequence prediction problem. Given a sequence of traffic measurements $M = \{M_1, M_2, ..., M_T\}$ where each $M_t \in \mathbb{R}^{H \times W}$ represents a spatial grid of $H \times W$ cells at time step $t$, our goal is to predict the next traffic volume grid $M_{T+1}$. Each grid cell contains a scalar value representing the traffic volume. This volume correlates with resource demands in the cellular network.


We define our input tensor as $X_t^{(i,j)} \in \mathbb{R}^{T \times K \times K}$, where $T$ is the sequence length (window of observation time steps), $K \times K$ represents the spatial dimensions of the input kernel, and $(i,j)$ are the spatial coordinate in the grid $M_t$. The prediction target is $x_{t+1}^{(i,j)} \in \mathbb{R}$,
representing the traffic volume at the center cell of the kernel for the next time step. Formally, we aim to learn a function $f: \mathbb{R}^{T \times K \times K} \rightarrow \mathbb{R}$ that minimizes the prediction error $\min_{\theta} L(f_{\theta}(X), x_{t+1}^{(i,j)})$, where $L$ is the loss function (i.e., Mean Absolute Error), and $\theta$ represents the learnable parameters of the model. The spatiotemporal nature of the data introduces a unique challenge: capturing both spatial correlations between neighboring cells and temporal dependencies across the sequence.

\subsection{Proposed Architecture: HiSTM}
We propose the Hierarchical SpatioTemporal Mamba (HiSTM), an architecture that combines hierarchical spatiotemporal processing with attention-based temporal aggregation  (Figure~\ref{fig:hstm}). Given an input tensor $\mathbf{X} \in \mathbb{R}^{T \times K \times K}$ (where $T$ is the number of time steps and $K \times K$ is the spatial kernel), the model predicts target values through three key components:

\begin{figure*}
    \centering
    \includegraphics[width=\textwidth]{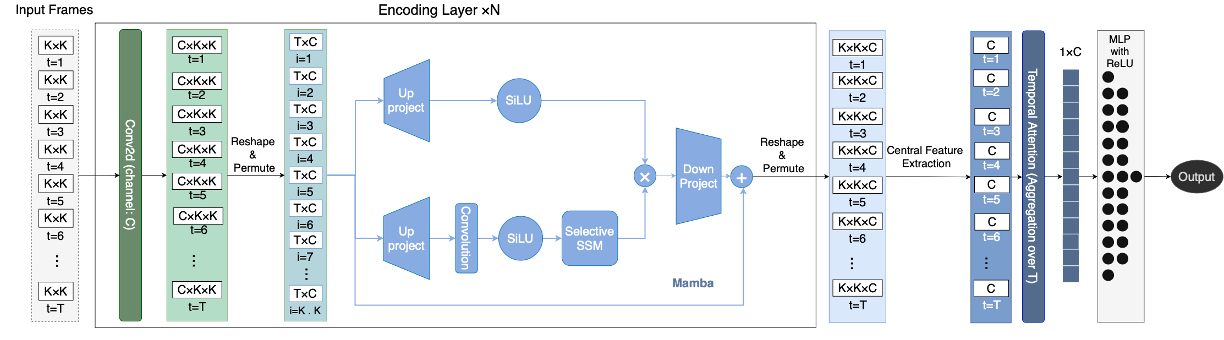}
    \caption{HiSTM Architecture}
    \label{fig:hstm}
\end{figure*}

\subsubsection{Hierarchical Spatiotemporal Encoding}
The input $\mathbf{X}$ is first augmented with an initial channel dimension (i.e., $D_{in}=1$), then passes through $N$ stacked Encoder Layers. Each Encoder Layer $l$ transforms its input $\mathbf{X}^{(l-1)}$ (where $\mathbf{X}^{(0)}$ is the initial augmented input) into an output $\mathbf{X}^{(l)} \in \mathbb{R}^{ T \times K \times K \times C}$. The operations within each layer are:
\begin{itemize}
    \item \textbf{Spatial Convolution}: Input features are first reshaped appropriately (e.g., to $ T \times D'_{in} \times K \times K$). A 2D convolution followed by a ReLU activation is applied. The first layer up-projects $D'_{in}$ to $C$ channels. Subsequent layers take $D'_{in}=C$ channels and output $C$ channels.
    \item \textbf{Temporal Mamba Processing}: The $C$-channel output from the convolution is reshaped to $\mathbf{X}_{\text{flat}} \in \mathbb{R}^{(K^2) \times T \times C}$. A Mamba SSM~\cite{gu2024mamba} then models the temporal dependencies for each of the $K^2$ spatial locations treated as sequences of length $T$, with $d_{\text{mamba}}=C$.
\end{itemize}

We use Mamba for its state space foundation, which models sequences through continuous-time dynamics rather than attention. This enables precise control over temporal structure and inductive bias, making it suitable for forecasting tasks where long-range temporal dependencies interact with fine-grained spatial patterns. Its ability to selectively retain and propagate information aligns with the demands of spatiotemporal modeling.

The Mamba output is reshaped back to $\mathbb{R}^{T \times K \times K \times C}$, forming $\mathbf{X}^{(l)}$. The output of this encoding stage is $\mathbf{X}_{\text{encoded}} = \mathbf{X}^{(N)}$.

\subsubsection{Temporal Attention-Based Aggregation}
From the encoded features $\mathbf{X}_{\text{encoded}} \in \mathbb{R}^{T \times K \times K \times C}$, features corresponding to the center spatial cell are extracted across all $T$ time steps. This yields a sequence $\mathbf{X}_{\text{center}} \in \mathbb{R}^{T \times C}$. An attention mechanism then computes an aggregated context vector $\mathbf{c} \in \mathbb{R}^{B \times C}$:
\begin{equation} \label{eq:temporal_attention}
    e_t = \text{Linear}_{\text{att}}(\mathbf{h}_t), \quad \alpha_t = \frac{\exp(e_t)}{\sum_{j=1}^T \exp(e_j)}, \quad \mathbf{c} = \sum_{t=1}^T \alpha_t \mathbf{h}_t
\end{equation}
where $\mathbf{h}_t \in \mathbb{R}^C$ is the feature vector from $\mathbf{X}_{\text{center}}$ (for a given batch instance) at time step $t$. The $\text{Linear}_{\text{att}}$ layer maps from $\mathbb{R}^C \to \mathbb{R}$, producing a scalar energy $e_t$. The softmax function normalizes these energies across all $T$ time steps to obtain attention weights $\alpha_t$.

\subsubsection{Prediction Head}
Finally, the aggregated context vector $\mathbf{c} \in \mathbb{R}^{C}$ is fed into a Multilayer Perceptron (MLP) head. This MLP consists of two linear layers with a ReLU activation in between, passing the dimensionality from $C$ to $MLP_{in}$, and then to 1, to produce the final prediction $\hat{y} \in \mathbb{R}^{1}$.

\subsection{Baseline Models}
We benchmark HiSTM against the following  baselines:

\begin{itemize}
\item \textbf{STN}~\cite{zhang2019long}: A deep neural network capturing spatiotemporal correlations for long-term mobile traffic forecasting.

\item \textbf{xLSTM}~\cite{xlstm}: A scalable LSTM variant with exponential gating and novel memory structures (sLSTM/mLSTM) designed as Transformer alternatives. For fair comparison, our implementation uses only one mLSTM layer (denoted as \texttt{xLSTM[1:0]} in their paper) to prioritize computational efficiency while retaining its parallelizable architecture.  
  
\item \textbf{STTRE}~\cite{sttre}: A Transformer-based architecture leveraging relative embeddings to model dependencies in multivariate time series.  
  
\item \textbf{VMRNN-B \& VMRNN-D}~\cite{vmrnn}: Vision Mamba-LSTM hybrids addressing CNN's limited receptive fields and ViT's computational costs. VMRNN-B (basic) and VMRNN-D (deep) use Mamba's selective state-space mechanisms for compact yet competitive performance.  
\end{itemize}

\section{Datasets \& Experimental Setup}

This section outlines the datasets, preprocessing steps, training configuration, and evaluation process used to assess model performance in spatiotemporal data forecasting.

\subsection{Datasets}
\begin{figure}
    \centering
    \includegraphics[width=\linewidth]{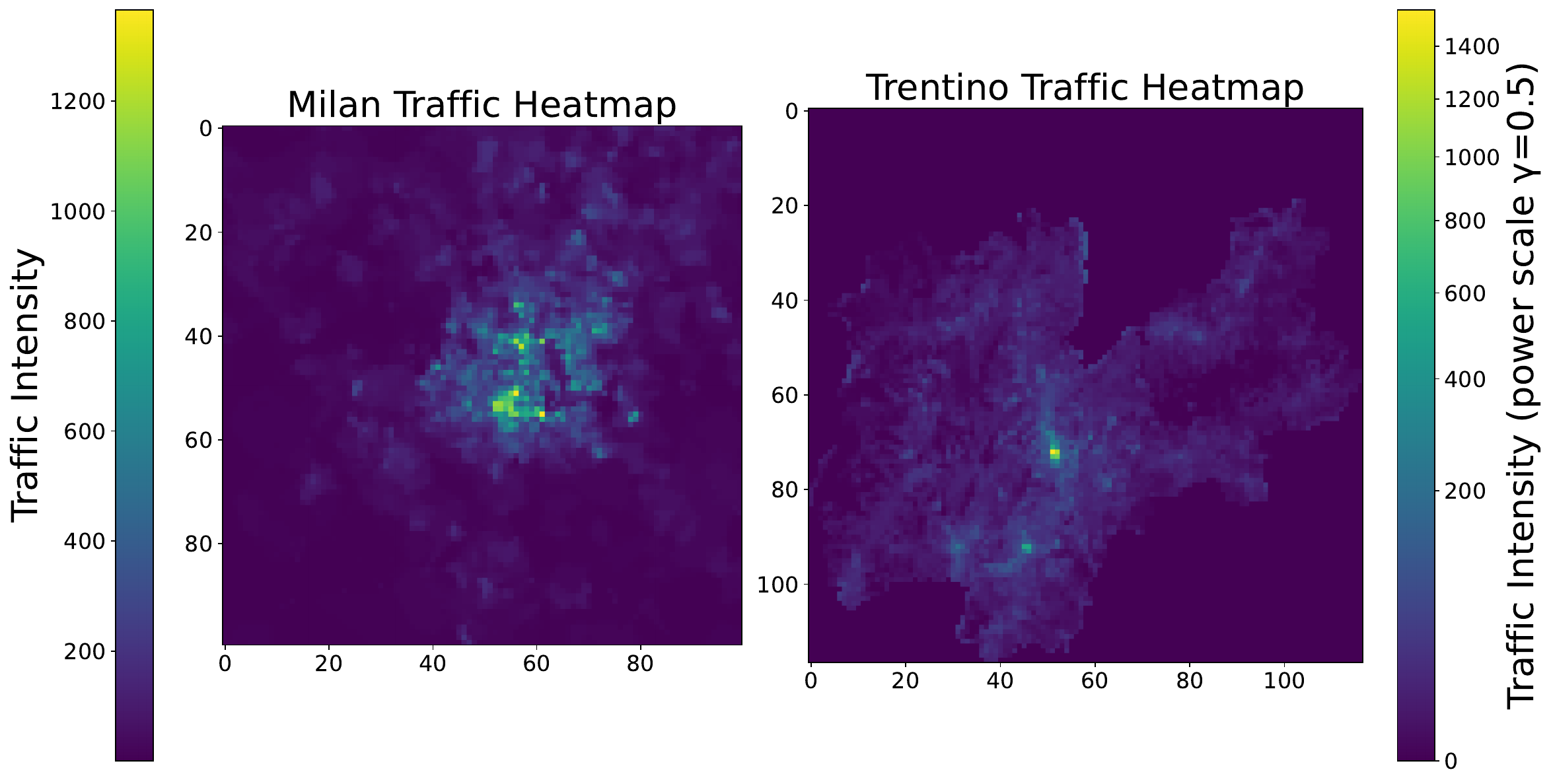}
    \caption{Spatial distribution of traffic flow intensity across Milan and Trentino regions.}    
    \label{fig:heatmap}
\end{figure}

We use the dataset introduced by Barlacchi et al.~\cite{datasets}, which contains two sub-datasets from Milan and Trentino with 10,000 and 11,466 spatial cells, respectively. Both datasets record SMS-in/out, Call-in/out, and Internet Traffic Activity in 10-minute intervals. The data exhibits spatial heterogeneity (see Figure~\ref{fig:heatmap}), with higher activity in central regions and localized clusters elsewhere, adding complexity to forecasting tasks. To better visualize these differences, the Milan heatmap uses a linear scale, while the Trentino heatmap employs a logarithmic scale to highlight the dense urban center against its sparsely populated surroundings. 

\begin{figure}
    \centerline{\includegraphics[width=8cm]{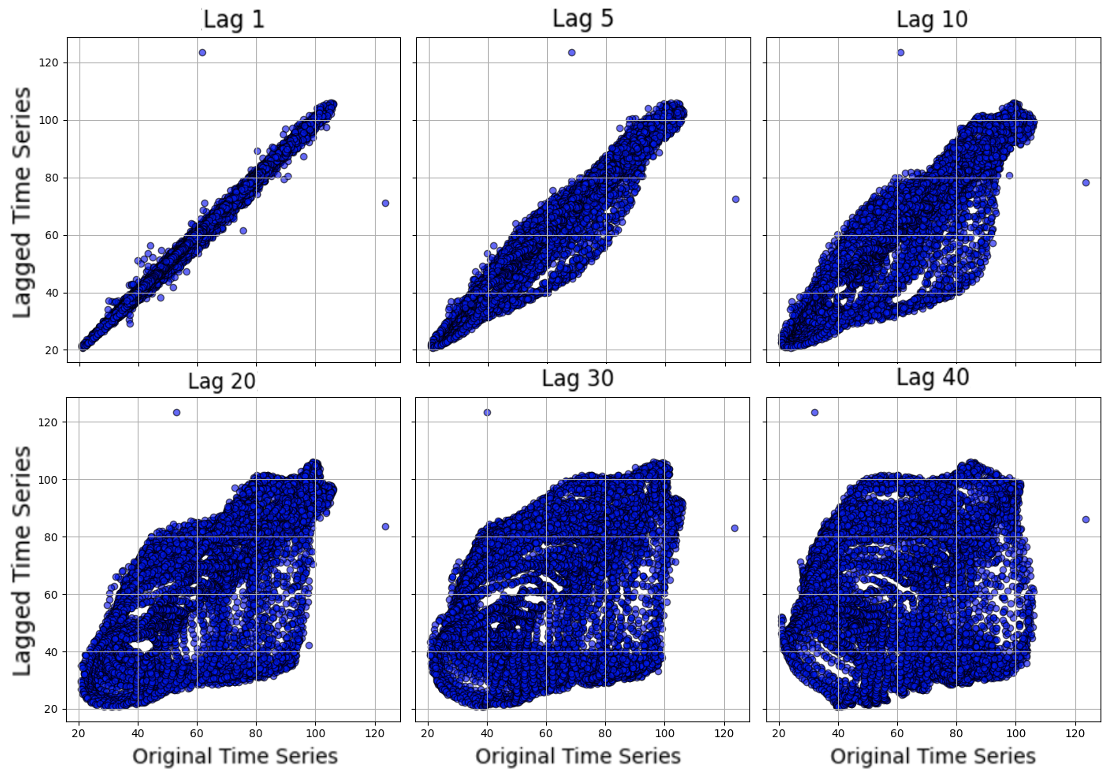}}
    \caption{Lag plot for the autocorrelation of the dataset for the entire grid.}\label{fig:lag_plot}
\end{figure}

Lag plot analysis (Figure~\ref{fig:lag_plot}) reveals that both datasets maintain correlation at higher lag values. Individual cell traffic demonstrates higher volatility (Approximate Entropy~\cite{pincus1991approximate} of 1.386 for a single cell) compared to spatially aggregated traffic (0.196), with the latter exhibiting enhanced cyclical patterns and passing the Augmented Dickey Fuller test~\cite{dickey1979distribution} for stationarity. It is evident that the aggregated series is more correlated with itself for different lags; hence, it is easier to predict. Consequently, incorporating the spatial element in the prediction can help the model to capture the distribution more effectively,  reduce the influence of individual events and magnify predictable cyclical patterns. This makes the series  more predictable in a spatiotemporal context compared to when using individual per-cell series.

\subsection{Data Preprocessing}
To ensure robust spatiotemporal feature extraction, chronological integrity, and leakage-free normalization for model training and evaluation, we preprocess the raw 100×100 spatial grid as follows. We set $K =11$, with boundary effects mitigated by cropping. Temporal sequences are constructed by aggregating six consecutive time steps. For the training set, input sequences are generated using a temporal stride of 6 between consecutive samples, minimizing temporal overlap from a single grid to promote feature diversity. During testing, a stride of 1 is used to ensure exhaustive evaluation across all temporal segments. The model is trained to predict the central value of the 11×11 kernel at the seventh (subsequent) time step. The preprocessed dataset is chronologically partitioned into training (70\%), validation (15\%), and test (15\%) sets to preserve temporal order and prevent information leakage from future to past.
Input features and target values are normalized to the \texttt{[0, 1]} range via Min-Max scaling, where scaling parameters are derived from the training data. Validation and test sets are transformed using these parameters, and out-of-range values are clipped to maintain the bounds.

\subsection{Implementation and Training Configuration}
We implement the HiSTM model in PyTorch, leveraging GPU-optimized operations for its \texttt{Mamba} and \texttt{Conv2D} modules to ensure computational efficiency. For reproducability, we release the complete source code for our HiSTM implementation in a public repository~\cite{repo}. For training and inference, we use an AI-server with a single NVIDIA A100 80GB GPU with 64 CPU cores and 512 GB RAM, using CUDA 12.4 and PyTorch 2.6.0+cu124. HiSTM and baseline models were trained with a batch size of 128 for up to 40 epochs, using early stopping with a patience of 15 and saving the best model based on validation loss. We use Adam optimizer with a learning rate of $10^{-4}$ and \texttt{ReduceLROnPlateau} scheduler (patience: 7, factor: 0.5).  

\subsection{Evaluation Metrics}
We evaluate prediction accuracy using four complementary metrics. Mean Absolute Error (MAE) measures the average absolute difference between predictions and ground truth. Root Mean Squared Error (RMSE) penalizes large errors more heavily. The coefficient of determination (R²) quantifies the proportion of variance explained by the model. Structural Similarity Index (SSIM) assesses the visual quality of spatial predictions. All metrics are computed after reversing the normalization to original scale.

\section{Results and Analysis}
This section evaluates forecast accuracy, cross-dataset generalization, and computational efficiency, comparing our proposed HiSTM model against baselines.

\subsection{Prediction Accuracy }

\subsubsection{Single-step Prediction Results}
HiSTM achieves the best single-step forecasting performance on the Milan dataset, with an MAE of 5.2196 and SSIM of 0.9925 (see Table~\ref{tab:single-step-milan}). Numbers are averaged across all spatial cells. This corresponds to a 29.4\% MAE reduction over STN and a 2.3\% SSIM gain, while outperforming other architectures such as STTRE (MAE: 5.5558) and parameter-intensive models like VMRNN-D (MAE: 6.4151). HiSTM also reports the lowest RMSE (11.2476) and the highest $R^2$ score (0.9799), indicating both lower large-error impact and improved variance explanation.

\begin{figure*}
    \centering
    \includegraphics[width=\linewidth]{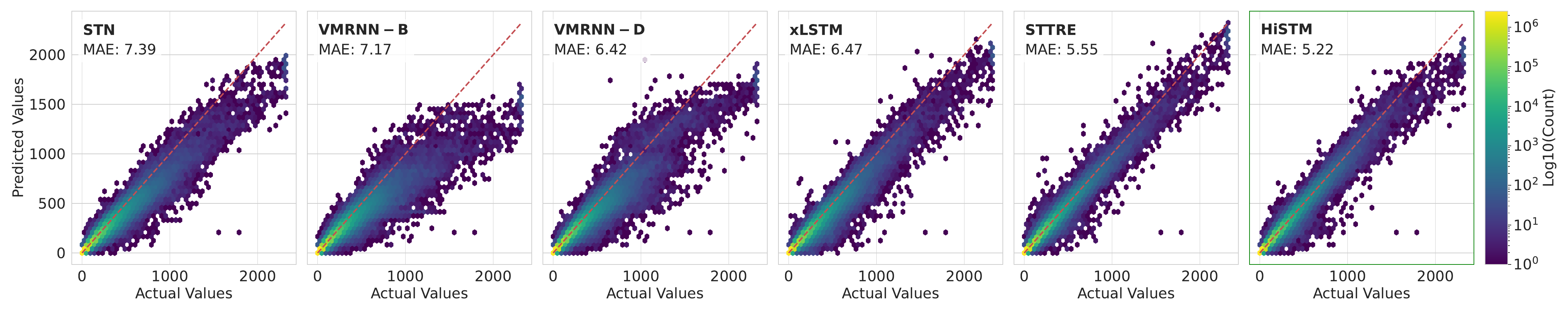}
    \caption{Density comparison of predicted vs. actual values over the entire grid for HiSTM and baseline models, with associated MAE}
    \label{fig:density}
\end{figure*}

In addition to numerical metrics, Figure~\ref{fig:density} provides a density-based comparison of predicted vs. actual values for all models. HiSTM shows a tighter concentration of points along the diagonal, particularly at higher traffic volumes, indicating improved predictive fidelity across the full dynamic range. It also exhibits fewer high-error outliers and lower dispersion compared to other models. STN and VMRNN-B, by contrast, display broader scatter, especially in the high-traffic regime, with more frequent large deviations. The distributional compactness in HiSTM suggests better generalization and robustness across heterogeneous spatial patterns.

\begin{table}
\centering
\caption{Single-step Prediction Performance on Milan Dataset. Best model indicated through bold font.}
\begin{tabular}{lcccc}
\toprule
\textbf{Model} & \textbf{MAE $\downarrow$}  & \textbf{RMSE $\downarrow$} & \textbf{R² Score $\uparrow$} & \textbf{SSIM $\uparrow$} \\
\midrule
STN & 7.3908 & 16.8824 & 0.9546 & 0.9853 \\
VMRNN-B & 7.1659 & 19.0876 & 0.9420 & 0.9843 \\
xLSTM & 6.4672 & 15.0901 & 0.9637 & 0.9870 \\
VMRNN-D & 6.4151 & 16.3284 & 0.9575 & 0.9873 \\
STTRE & 5.5558 & 11.4426 & 0.9791 & 0.9917 \\
HiSTM & \textbf{5.2196} & \textbf{11.2476} & \textbf{0.9799} & \textbf{0.9925} \\
\bottomrule
\label{tab:single-step-milan}
\end{tabular}
\end{table}

\subsubsection{Multi-step Autoregressive Forecasting}
HiSTM demonstrates improved stability over extended forecasting horizons (Table~\ref{tab:multi-step-milan}). While all models accumulate error with each additional step, HiSTM maintains the lowest MAE and RMSE across all six steps. At step 6, HiSTM reports an MAE of 6.69, which is 36.8\% lower than STN (10.59) and 11.3\% below STTRE (7.54). The slope of MAE progression is 58\% lower than that of STN, indicating reduced error propagation. SSIM also degrades more gradually in HiSTM, remaining above 0.95 even at step 6. This suggests that HiSTM better preserves temporal dependencies and structural consistency across time.

\begin{table*}
\centering
\caption{Performance comparison of models over multiple steps (autoregressive forecasting). Best model indicated through bold font.}
\resizebox{\textwidth}{!}{%
\begin{tabular}{c|ccc|ccc|ccc|ccc|ccc|ccc}
\toprule
\multirow{2}{*}{Step} & \multicolumn{3}{c|}{HiSTM} & \multicolumn{3}{c|}{STN} & \multicolumn{3}{c|}{STTRE} & \multicolumn{3}{c|}{xLSTM} & \multicolumn{3}{c|}{VMRNN-B} & \multicolumn{3}{c}{VMRNN-D} \\
    & MAE & RMSE & SSIM & MAE & RMSE & SSIM & MAE & RMSE & SSIM & MAE & RMSE & SSIM & MAE & RMSE & SSIM & MAE & RMSE & SSIM \\
\midrule
1 & \textbf{3.87} & \textbf{9.54} & \textbf{0.9833} & 5.28 & 13.05 & 0.9742 & 4.21 & 9.83 & 0.9808 & 4.64 & 11.86 & 0.9759 & 5.11 & 13.75 & 0.9745 & 4.62 & 12.25 & 0.9777 \\
2 & \textbf{4.38} & \textbf{10.95} & \textbf{0.9785} & 6.44 & 16.29 & 0.9615 & 4.88 & 11.45 & 0.9739 & 5.62 & 14.63 & 0.9639 & 5.08 & 13.85 & 0.9740 & 4.61 & 12.36 & 0.9772 \\
3 & \textbf{4.85} & \textbf{12.06} & \textbf{0.9735} & 7.39 & 18.54 & 0.9490 & 5.47 & 12.73 & 0.9668 & 6.47 & 16.78 & 0.9509 & 6.06 & 16.33 & 0.9637 & 5.35 & 14.30 & 0.9698 \\
4 & \textbf{5.56} & \textbf{13.42} & \textbf{0.9680} & 8.59 & 21.05 & 0.9335 & 6.27 & 14.12 & 0.9586 & 7.50 & 18.95 & 0.9356 & 6.83 & 17.64 & 0.9561 & 6.08 & 15.49 & 0.9641 \\
5 & \textbf{6.09} & \textbf{14.62} & \textbf{0.9633} & 9.55 & 22.94 & 0.9196 & 6.88 & 15.28 & 0.9513 & 8.34 & 20.58 & 0.9217 & 7.45 & 18.91 & 0.9492 & 6.59 & 16.54 & 0.9591 \\
6 & \textbf{6.69} & \textbf{16.02} & \textbf{0.9578} & 10.59 & 25.01 & 0.9033 & 7.54 & 16.60 & 0.9430 & 9.23 & 22.38 & 0.9064 & 8.24 & 20.38 & 0.9398 & 7.30 & 17.87 & 0.9520 \\
\bottomrule
\end{tabular}%
}
\label{tab:multi-step-milan}
\end{table*}

\subsubsection{Cross-dataset Generalization on Trentino dataset}
On the unseen Trentino dataset, HiSTM achieves a 47.3\% MAE reduction (1.3870 vs. 2.6344) and a 36.9\% RMSE improvement (4.8134 vs. 7.6370) over STN (see Table~\ref{tab:single-step-trentino}). It also yields the highest SSIM (0.9916) and $R^2$ score (0.9649), confirming strong structural preservation. HiSTM outperforms all baselines across all evaluation metrics, including STTRE (MAE: 1.8132), VMRNN-D (MAE: 1.5870), and VMRNN-B (MAE: 1.9751). This demonstrates HiSTM's capacity to generalize to new spatial environments with distinct activity distributions and scales.

\begin{table}
\centering
\caption{Single-step Generalization Performance on Trentino Dataset. Best model indicated through bold font.}
\begin{tabular}{lcccc}
\toprule
\textbf{Model} & \textbf{MAE $\downarrow$} & \textbf{RMSE $\downarrow$} & \textbf{R² Score $\uparrow$} & \textbf{SSIM $\uparrow$} \\
\midrule
STN & 2.6344 & 7.6370 & 0.9116 & 0.9762 \\
xLSTM & 2.5974 & 8.9235 & 0.8793 & 0.9615 \\
VMRNN-B & 1.9751 & 6.6270 & 0.9334 & 0.9839 \\
STTRE & 1.8132 & 5.0050 & 0.9620 & 0.9903 \\
VMRNN-D & 1.5870 & 5.5754 & 0.9529 & 0.9885 \\
HiSTM & \textbf{1.3870} & \textbf{4.8134} & \textbf{0.9649} & \textbf{0.9916} \\
\bottomrule
\label{tab:single-step-trentino}
\end{tabular}
\end{table}

\subsection{Cell-specific Modeling and Spatially-aware Accuracy}
\begin{figure}
    \centering
    \includegraphics[width=\linewidth]{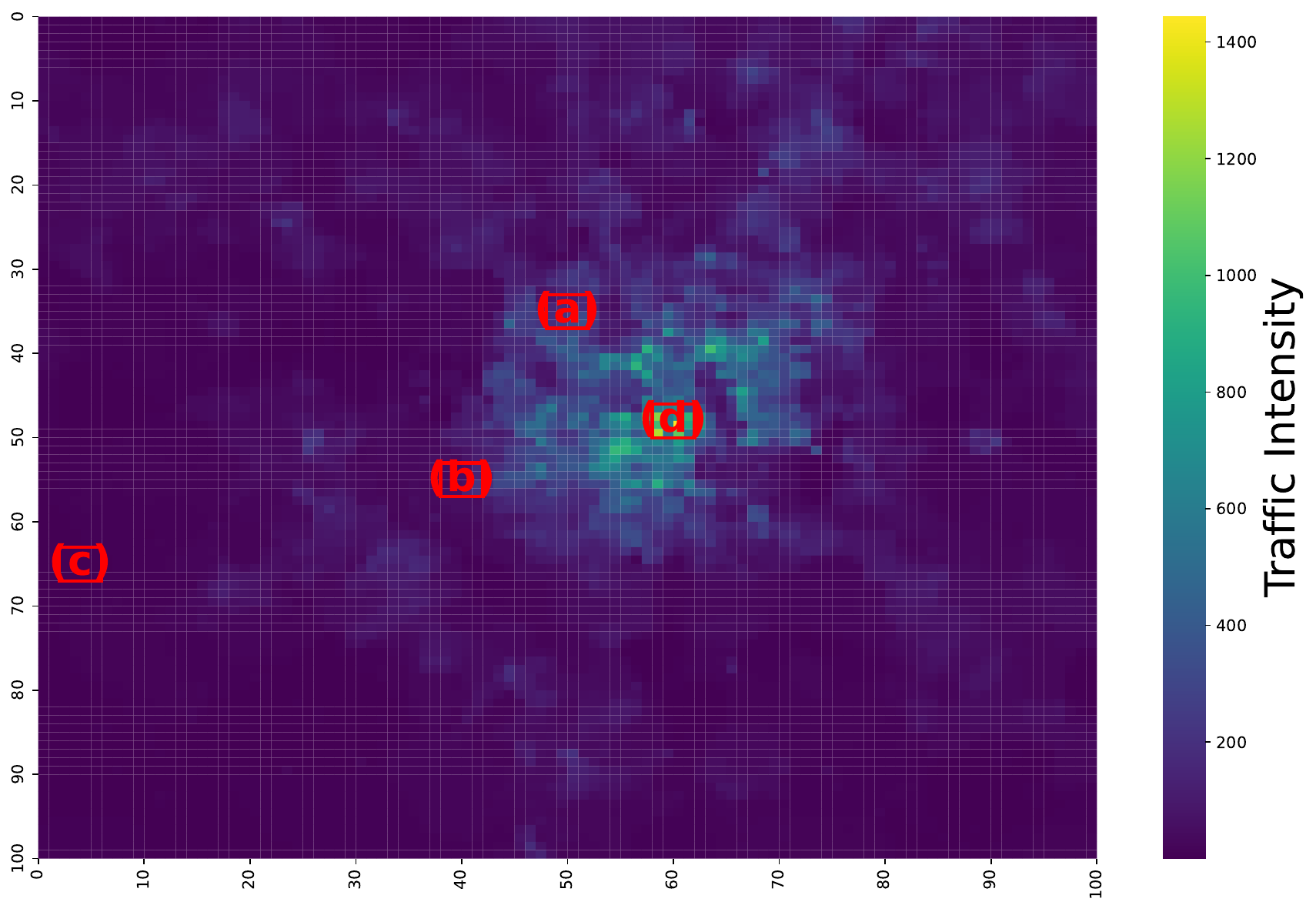}
    \caption{Selected cells from the Milan's traffic network. The cells represent different traffic patterns: (a) urban, (b) suburban, (c) rural, and (d) maximum variance cell.}    
    \label{fig:analysis-locations}
\end{figure}

\begin{figure}
    \centering
    \includegraphics[width=\linewidth]{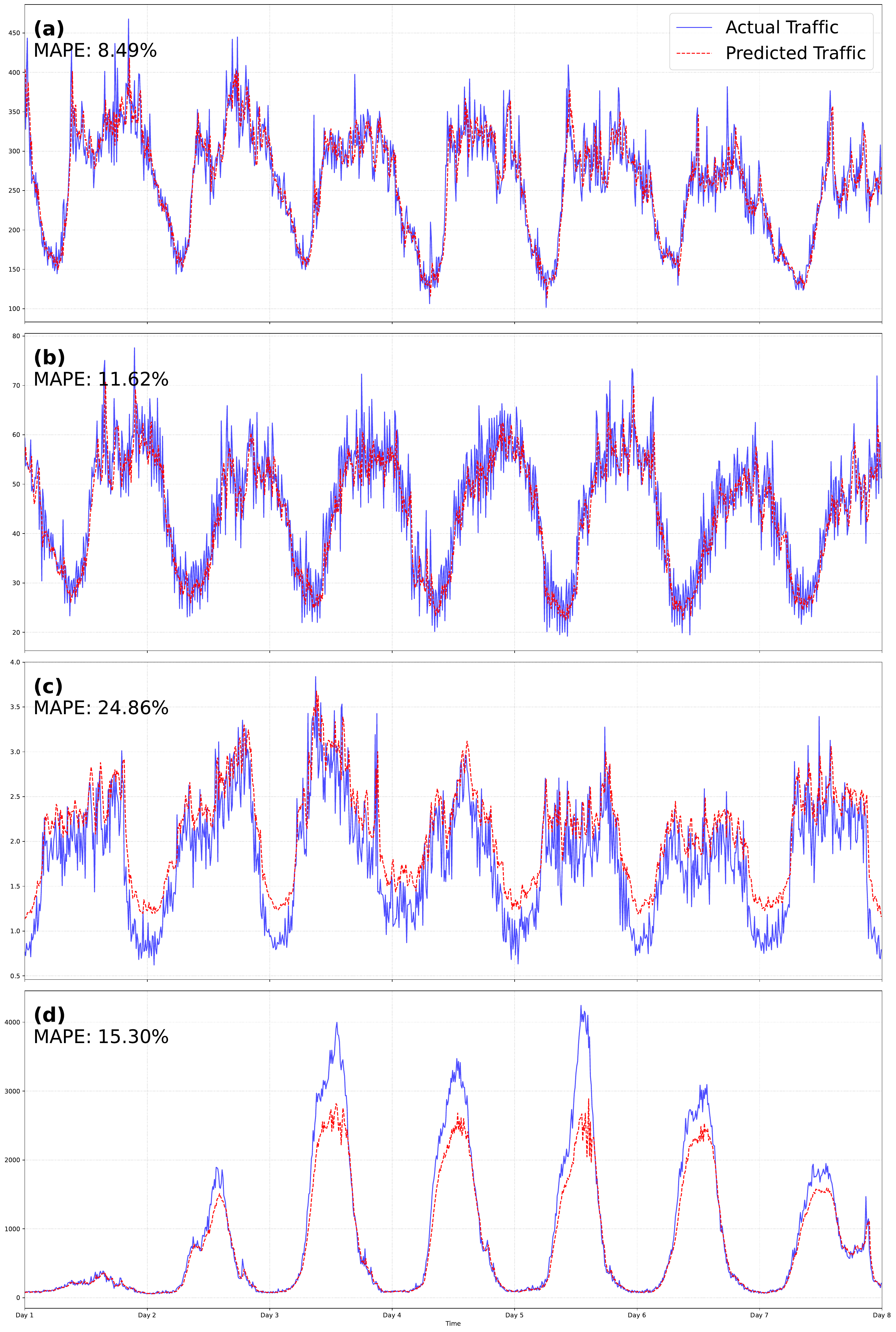}
    \caption{Week-long traffic predictions from the HiSTM model across four representative cells in Milan's traffic network. The cells were selected to represent different traffic patterns: (a) urban, (b) suburban, (c) rural, and (d) maximum variance cell. Time is shown in days, with each day containing 144 readings (10-minute intervals)}    
    \label{fig:analysis}
\end{figure}

To evaluate the spatially-aware accuracy of the model, we use Milan dataset and select a random 7-day time period (1008 time steps) from the test dataset. We use a 6-step memory  window and an 11-step kernel on a 100×100 sensor grid to forecast a single next timestep. To analyze spatial performance across varying traffic conditions, we select four representative cells, corresponding to (a) urban (b) suburban (c) rural and (d) the cell with the maximum temporal variance (Figure~\ref{fig:analysis-locations}). The predicted versus actual traffic trajectories for all four cells are shown in Figure~\ref{fig:analysis}.
 
The model achieved its lowest MAPE (8.49\%) for the urban cell, demonstrating strong performance in dense traffic zones. Prediction accuracy remained stable for the suburban cell (11.62\% MAPE) with only minor deviations during peak periods. The rural cell showed higher relative error (24.86\% MAPE), where low absolute traffic volumes magnified percentage errors. Notably, the model maintained robust performance (15.30\% MAPE) for the high-variance cell despite its extreme fluctuations, highlighting HiSTM's capacity to handle volatile temporal patterns while showing expected limitations in sparse rural conditions where signal-to-noise ratios are challenging. These performance differences reflect the interaction between model complexity and regional traffic characteristics: urban and suburban areas offer rich temporal patterns and denser signals that align well with the model’s spatiotemporal structure, whereas rural areas pose challenges due to sparse activity and fewer recurring patterns. The high-variance cell underscores the model's ability to generalize to non-stationary behavior, though peak underestimation suggests further headroom for improvement in handling extremes.

\subsection{Computational Efficiency Analysis}

HiSTM achieves a better performance while requiring fewer model parameters  (Table~\ref{tab:model-comparison}). At 33.8K parameters (0.13 MB), it requires 18× fewer parameters than VMRNN-D and 5.4× less than xLSTM while delivering faster inference (1.19 ms) than all baselines, except xLSTM. Though its multiply-accumulate operations (MACs, a standard metric for computational workload) total $1.36\times10^{7}$—higher than STN's—HiSTM balances computational cost with accuracy, achieving a MAC/MAE ratio 2.6× better than STTRE’s. This efficiency-profile positions HiSTM as a practical solution for resource-constrained deployment scenarios.

\begin{table}
\centering
\caption{Model Comparison}
\begin{tabular}{lccccc}
\toprule
\textbf{Model} & \makecell{\textbf{Parameter} \\ \textbf{Count}} & \makecell{\textbf{Size} \\ \textbf{(MB)}} & \makecell{\textbf{GPU} \\ \textbf{(MB)}} & \makecell{\textbf{Inference} \\ \textbf{(ms)}} & \textbf{MACs} \\
\midrule
xLSTM & 607,753 & 2.32 & 13.69 & 1.01 & \textbf{$5.96 \times 10^5$} \\
VMRNN-B & 137,282 & 0.52 & \textbf{9.77} & 8.16 & $2.06 \times 10^7$ \\
VMRNN-D & 1,506,498 & 5.75 & 15.47 & 18.58 & $5.52 \times 10^7$ \\
STTRE & 165,380 & 0.63 & 58.07 & 4.54 & $2.83 \times 10^7$ \\
STN & 576,755 & 2.20 & 11.34 & 2.46 & $2.31 \times 10^6$ \\
HiSTM & \textbf{33,794} & \textbf{0.13} & 10.63 & \textbf{1.19} & $1.36 \times 10^7$ \\
\bottomrule
\label{tab:model-comparison}
\end{tabular}
\end{table}

\section{Conclusion}\label{conclusion}

In this paper, we presented HiSTM, a hierarchical spatiotemporal model for efficient and accurate cellular traffic forecasting. By combining dual spatial encoders with a Mamba-based temporal module and an attention mechanism, HiSTM captures complex spatiotemporal patterns with minimal overhead. Experiments on real-world datasets show that HiSTM demonstrates competitive performance, reducing MAE by 29.4\% compared to STN baseline while maintaining computational efficiency. It generalizes well to the unseen Trentino dataset and sustains lower errors over longer forecast horizons. These results suggest HiSTM's potential as an efficient approach for cellular traffic prediction, though broader validation across diverse network conditions would strengthen deployment recommendations.

Future work will explore mixture-of-experts approaches to better model spatially-clustered traffic patterns and kernel-to-kernel forecasting to capture finer-grained temporal dynamics across service types. We aim to investigate diffusion-based decoding strategies to enhance long-range predictive capabilities and extend the model's scope through different aggregation strategies.

To strengthen generalizability, we will evaluate HiSTM on diverse geographical datasets beyond Italy and implement attention visualization for improved interpretability. We also plan to address practical deployment challenges including missing data handling, concept drift adaptation, and spatial heterogeneity through adaptive weighting mechanisms.

\section*{Acknowledgement}
This work was partly funded by the Bavarian Government by the Ministry of Science and Art through the HighTech Agenda (HTA).

\bibliographystyle{IEEEtran}

\end{document}